\documentclass{article}   
\usepackage{natbib}

\usepackage{physics,amsmath,amssymb,mathrsfs}
\usepackage{enumitem,appendix}

\title{Empty space and the (positive) cosmological constant}
\author{Mike D. Schneider\footnote{Department of Philosophy, University of Missouri} \footnote{Work on this article began nearly six years ago, and the project was intermittently shelved on several occasions since. As a result, I am sorry if I have forgotten anyone who I had intended to thank throughout the process. In any case, the names that come readily to my mind are: Jim Weatherall, Gordon Belot, John Earman, and David Wallace. I am also grateful to my audience at the Center for Philosophy of Science at University of Pittsburgh two years ago when, during my postdoc there, I rekindled interest in furthering the project. Finally, I appreciate the patience of my anonymous reviewers --- I am sure my main arguments are stronger and clearer owing to their help.}}
\date{March 2023}

\begin{document}
\maketitle

\begin{abstract}
I discuss empty space, as it appears in the physical foundations of relativistic field theories and in the semiclassical study of isolated systems. Of particular interest is the relationship between empirical measurements of the cosmological constant and the question of appropriate representation of empty space by spacetimes, or models of general relativity. Also considered is a speculative move that shows up in one corner of quantum gravity research. In pursuit of holographic quantum cosmology given a positive cosmological constant, there is evidently some freedom available for theoretical physicists to pick between two physically inequivalent spacetime representations of empty space, moving forward: de Sitter spacetime or its `elliptic' cousin.
\end{abstract}
%keywords: cosmological constant; de Sitter spacetime; speculation in science

\section{Introduction}

Since the 1990s, standard model cosmology has included a positive-valued `cosmological constant'. Its role is that of a fit parameter to accommodate data about the slowed growth rate of large-scale structure within our mean expanding cosmos. My first goal in this article is to provide a principled argument in support of a claim in the foundations of fundamental physics, which is often entertained implicitly and concurrently with these otherwise remote empirical developments. 

The claim is that we ought (in light of such empirical developments) to seriously pursue theories of matter in fundamental physics, which in the absence of gravity take for granted spacetime structure locally equivalent to de Sitter spacetime (hereafter `dS') and globally quite restricted. Implicit invocations of this claim within the fundamental physics literature are too numerous to track. Examples are to be found almost wherever projects take on technical challenges in quantum field theory or effective field theory that arise when the cosmological constant is allowed to be fixed as something positive, on the basis of empirical claims.\footnote{By this criteria, examples in mind inevitably come in close contact with the `cosmological constant problem'; \citep{bianchi2010all} provide a polemic on the latter, from a perspective that echoes the one introduced here. Setting aside the expansive literature within quantum field theory on curved spacetime specifically preoccupied with formulating field theories on dS given the importance of dS to cosmology \citep{akhmedov2014lecture}, other examples include work in introducing a matter sector within loop quantum gravity  \citep{smolin2002quantum,dupuis2014observables}, as well as analogous work on formulating `de Sitter' low-energy quantum gravity in a string theory or holographic approach \citep{banks2000cosmological,kachru2003sitter,banks2006towards}. Concerning work in the latter, persistent difficulties encountered over the past two decades have led to `swampland' conjectures that the string theory landscape may exclude all low-energy effective field theories with de Sitter or de Sitter-like backgrounds \citep{palti2019swampland}. Some have argued that support for these conjectures challenges the overall promise of string theory, exactly because of the empirical successes of both our current theories of matter and standard model cosmology \citep{danielsson2018if,silk2022swampland}.} Of course, in many such instances, the core contribution of the work may very well be decoupled from the implicit claim. But it is my view that, in such instances, the onus is rather placed on those who would wish to do so.

To my knowledge, this claim --- so often left implicit, anyway --- is nowhere explicitly defended. But it clearly has something to do with a longstanding view in general relativistic cosmology that non-zero, signed values of the cosmological constant in our understanding of cosmic expansion are indicative of a new fundamental constant of nature: compare, e.g., old remarks by Tolman to Einstein in 1931 reproduced in \citep[p. 197]{earman2001lambda} with the recent concluding remarks of \citet{silk2022swampland}. 

In fact, more than just that longstanding view would be needed in order to substantiate the claim. As I will suggest below, the complete argument in support of the claim is best construed as a novel thesis about appropriate spacetime representations of \emph{empty space} relevant in the foundations of relativistic field theories, given general relativity as our current best, fully non-perturbative theory of gravity. And the conclusion of that argument bears immediately as well on foundational work in asymptotics within the perturbative, field-theoretic study of isolated semiclassical gravitational systems, given measurements of a positive cosmological constant. The argument is thereby well suited to make physical sense of the research program in dS asymptotics inaugarated by \citet{ashtekar2014asymptotics}, in light of the above empirical developments in standard model cosmology. 

Postponing general foundational discussion about dS asymptotics, as well as about reformulating various familiar matter theories absent gravity `in the case of a positive cosmological constant', my second goal in the article is instead to pursue just one (other) intriguing upshot of the novel thesis --- now in the context of quantum gravity. Namely, one consequence of the thesis is that the global spacetime structure apt for empty space in the case of a positive cosmological constant fails to be uniquely specified among two live options. dS appropriately represents empty space, but so does its topological cousin, elliptic dS (hereafter `EdS'). As I will argue, the identification of an equivalence class of spacetimes consisting of dS and EdS as the appropriate spacetime representation of empty space given general relativity does well to contextualize certain speculative projects in holographic quantum cosmology. In particular, given a positive cosmological constant, one often sees in that research a freedom asserted: to swap out dS for EdS, moving forward. The analysis I will provide of this asserted freedom based on the thesis about empty space ultimately demonstrates an important lesson in the epistemology of theoretical research in science. Although theoretical research might appear to be a radically free and unconstrained intellectual activity (namely: by dint of the ineliminable role of speculation within), what one sees in this particular asserted freedom is a certain form of continuity in speculation itself.  The relevant speculation rests firmly on a very particular conclusion within the foundations of our contemporary physics --- something I will later suggest is an instance of open texture within current theory, which leaves room for precisely one such speculation.

That conclusion, as I have already indicated, concerns the appropriate representation of `empty space' in relativistic field theories. But framing discussion in terms of `empty space' might be misleading, as there is no canonical understanding of the term. As such, I turn now to clarify what I mean.

\section{Empty space}

`Empty space' is my own term. What it designates is the arena in our current understanding of our current best physics, within which one may pursue any low-energy study of matter --- and this, specifically while claiming to be in a sector of the relevant matter theory, which is characterized by the noted absence of any gravitational coupling or interaction (except, perhaps, as may then be introduced in space and time according to a relativistic perturbation theory). But getting analytic traction on the concept of empty space is secondary to getting clear on its appropriate representation in relativistic field theories --- the only relevant low-energy matter theories (classical or pre-quantum\footnote{That is, here I take the view that (relativistic) quantum field theories, like those familiar in standard model particle physics, may be understood as governing quantized relativistic fields, situated on a spacetime.}) we currently have. 

This last aim is accomplished by answering a question:
\begin{itemize}
\item[(Q)] In the wake of general relativity, what is the relevant spacetime background for relativistic field theories in the low-energy study of matter, when considering such material systems as may be characterized by a complete lack of gravity?
\end{itemize}
Though I have never seen Q articulated as such, it strikes me as immediately motivated. First, it is well understood that relativistic field theories, as our current best theories of matter, require some or other choice of spacetime background: a `target' relativistic spacetime on which we may define the relevant field in question as something `relativistic', yet which is also situated spatiotemporally \citep[\S 2]{fletcher2022local}. Meanwhile, equally well understood is that the mathematical models of general relativity happen to be Lorentzian manifolds, physically interpreted as relativistic spacetimes. Moreover, general relativity is important for our contemporary understanding of gravity: accordingly, descriptions of the possible gravitational states of self-gravitating systems, e.g. that of our total universe, are rendered fully non-perturbatively in terms of relativistic spacetime geometry. So, it would seem that the spacetime background relevant in the context of relativistic field theories, in the wake of general relativity, inevitably invites us to say \emph{something} about the gravitational state of the system whose physics we might otherwise have thought to be exhausted by descriptions of the relativistic field situated on top.

And yet, second, it is standard to claim relativistic field theories as crucial tools in contemporary physics of relevance in the low-energy study of matter, \emph{specifically when ignoring any ambient influence of gravity}. Here I have in mind especially the common claim that standard model particle physics captures everything in fundamental physics \emph{but} gravity. Consequently, it seems that there is some further reason to ask and answer Q, in light of standard practice. Namely: one would like some means of understanding matter --- in the context of the employ of relativistic field theories, which make use of spacetimes in the background --- as \emph{nonetheless} quarantined from a concurrent general relativistic understanding of gravity. 

Through the next two sections, I will develop one plausible answer to Q. This is an answer that turns out to be suitable to physically motivate the program of research in dS asymptotics inaugurated by \citet{ashtekar2014asymptotics} (see also  \citep{ashtekar2017implications}): where, given a positive cosmological constant, isolated systems are conceived perturbatively in terms of localized physics situated at a spatial origin in an asymptotically de Sitter setting, rather than in an asymptotically Minkowski setting that is reserved for the case of a vanishing cosmological constant. In this program, although researchers are foremost focused on how to adequately model semiclassical (non-perturbative) gravity by means of perturbative methods, one might also be concerned more generally with the continued reliance in particle physics on various relevant energy and particle concepts, like ADM mass, charge, and angular momentum, which take for granted an ideal perturbative description of isolated systems that includes asymptotic features of Minkowski spacetime. A reliance on those features of Minkowski spacetime is particularly troubling, given that Minkowski spacetime would seem to occupy a singular `$\text{cosmological constant} \rightarrow 0$' limit of de Sitter spacetimes of diminishing curvature.\footnote{dS is a maximally symmetric spacetime with constant positive curvature fully characterized by the value of the positive cosmological constant (more on this in section \ref{secbreaking}). Minkowski spacetime is the maximally symmetric flat spacetime, i.e. with constant $0$ curvature. So it is natural to consider Minkowski spacetime as locally approximating dS in the theory of possible spacetime backgrounds for isolated systems modeled locally at a spatial origin in terms of relativistic matter fields. But globally, Minkowski spacetime and dS fail to be diffeomorphic. As a result, there is a natural sense in which the limit is singular for a sequence of de Sitter spacetimes of diminishing constant curvature, which approaches $0$ curvature: the closed spatial dimensions of dS are, in the limit, abruptly torn asunder. On one tradition in philosophy of physics, singular limits like this one are prima facie concerning, particularly regarding matters of intertheory reduction, approximation, and understanding \citep{berry1995asymptotics,batterman2001devil,bokulich2008reexamining}. It is therefore reasonable to ask whether there are conceptually satisfying ways of recasting the same limit sequences as non-singular, i.e. as continuous (see \citep{steeger2021classical} for general discussion). Notably, in the context of certain sequences of black hole spacetimes of diminishing ADM mass, \citet{geroch1969limits} proposed an approach to taking spacetime limits globally that is also relevant here. The approach stitches together frames imposed on each spacetime in a given sequence, in effect bypassing the need for the spacetimes to be diffeomorphic in order for the sequence to be regarded as continuous. This approach can be applied to the present case, for instance using the static patch coordinates on dS defined about an arbitrary maximal timelike geodesic. The catch is that, to thereby interpret the limit as non-singular, one must interpret the stitching of the frames as itself physically significant, for instance (in the case of the static patch frame) representing a continuous rescaling of dimensionful units as one moves through the sequence. Unfortunately, in addition to being highly frame-specific, this physical interpretation would seem less than conceptually satisfying: the specified frame in the present case, which one might otherwise interpret as endowing spacetime with a system of dimensionful units, is observer-relative and fails to be globally well defined.} (Of course, this is only troubling in the first place, provided there is an argument linking a positive cosmological constant to a foundational turn away from Minkowski spacetime and toward dS --- precisely what my answer to Q will provide.\footnote{\label{fndeflationary}A reviewer points out that particle physics would seem to get on just fine with theories defined on Minkowski spacetime, in their analyses of scattering experiments. But since every relativistic metric approximates every other at a point \citep{fletcher2022local1}, it should not be surprising to find that in the vicinity of a collision event, there exists an empirically adequate description of that collision provided by theories defined on just one particular spacetime, which might just as well --- for convenience --- be Minkowski spacetime. (Convenience isn't everything; \citet{koberinskiSmeenkRethinkingCCP} critically discuss the role of global spacetime structure, e.g. as would be provided by Minkowski spacetime in this case, in formulating effective field theories in particle physics. See also \citep[\S 3]{koberinski2021regularizing}.) The worry, meanwhile, is that our thinking in particle theory is distorted by global features of the particular spacetime chosen by convenience for these purposes, and those global features are clearly themselves beyond the empirical scope of the scattering experiments. Implicit in this worry is a belief, motivated by the history of work on black holes (see also the previous footnote), that we ought not by default be deflationary about global differences in spacetime structure, beyond local differences as are manifestly relevant to descriptions of the conditions within such experiments.})

So, with a program like dS asymptotics in the back of the mind, consider now three talking points about our contemporary physics, which I claim together implicitly serve as a surrogate to Q's answer. These talking points are, first, that by a tradition hearkening back to the identification of the lightcone structure within special relativity, Minkowski spacetime stands as the eminently natural choice of appropriate representation of empty space in relativistic field theories; Minkowski spacetime is the historically default kinematic environment when ignoring the influence of gravity in our relativistic universe, and by extension asymptotic features of Minkowski spacetime provide a default in defining a relativistic, field-theoretic perturbation theory suited for isolated systems. Second, \emph{seemingly any empirically viable precision measurement} of a cosmological constant in general relativity, e.g. via standard model cosmology, heralds just cause to break from that tradition. And third, on the break from that tradition, Minkowski spacetime exits the scene and dS (along with EdS --- more on this in section \ref{secbreaking}) enters instead. Filling out in sections 3 and 4 how these three talking points can in fact hang together as a well founded answer to Q will put me in a position to motivate the research program in dS asymptotics. It will also enable me to substantiate, in section 5, my analysis sketched in the Introduction on the interpretation of the asserted freedom in holographic quantum cosmology to swap between dS and EdS. 

\section{A traditional view}

The first of the three talking points invokes a notion of `tradition', which is ultimately broken by some means (i.e. the second talking point) to particular consequences for our current understanding of our current best physics (i.e. the third talking point). In this section, I develop the tradition I have in mind.

General relativity is a theory of relativistic spacetime geometry.\footnote{Standard references include \citep{malament2012topics}, \citep{wald1984}, and \citep{hawking1973large}. In what follows, I will take for granted four spacetime dimensions, though the theory readily admits generalizations of interest in quantum gravity research.} The central equation within the theory provides a \emph{geometro}dynamics for classical gravity: the equation describes a dynamical coupling between spacetime geometry and the distribution of stress-energy associated with all classically gravitating matter. Vacua, or spacetime geometries that locally satisfy a `source-free' sector of the geometrodynamics, are then merely spacetimes that happen to be devoid of dynamically relevant stress-energy, i.e. gravitating matter.

Clearly, which spacetimes count as vacua within general relativity is sensitive to the exact form of the source-free sector of the geometrodynamics. But the exact form of that central equation within the theory has been something of a minor controversy throughout the entire history of the theory's development \citep{earman2001lambda}. As has been substantiated, rather influentially, by \citet{lovelock1971einstein} as a matter of tensorial mathematics on fixed arbitrary manifolds, about the only point of agreement throughout history is that the central equation within general relativity is a member within a whole family of equations --- the `Einstein equations' --- that are naturally associated with the affine structure $\mathbb{R}$. Each indexing of the family by $\mathbb{R}$ that respects that affine structure can then be interpreted as picking out a different distinguished family member: whichever equation is accordingly in the image of $0$. 

This is to speak of the possibility of a new Real-valued fundamental constant of nature, which is motivated by philosophical reflection on the underlying mathematical structure of general relativity, and which is to be incorporated into any exact specification of the gravitational theory (though, see \citep[\S3]{earman2022trace} for a caveat). While it is standard to consider the explicit indexing that takes values $\lambda\in\mathbb{R}$ to the `Einstein equation with cosmological constant $\lambda$' --- so that the distinguished family member is the `Einstein equation with cosmological constant $0$' (often shortened as `Einstein's equation') --- one may also consider explicit indexings that take values $\lambda\in\mathbb{R}$ to the `Einstein equation with cosmological constant $\Lambda$', where $\Lambda=\lambda+\lambda_0$ for some $\lambda_0\in\mathbb{R}$ that is often dubbed the `bare' cosmological constant. In this case,  the distinguished family member is the `Einstein equation with cosmological constant $\lambda_0$', or just `Einstein's equation with a cosmological constant' (the fixed value of the bare term $\lambda_0$ usually being left implicit).\footnote{Note that, as defined, `Einstein's equation' is a special case of `Einstein's equation with a cosmological constant'. Just so: it is Einstein's equation with a cosmological constant, whenever the implicit bare term is decided as $0$ --- i.e. as opposed to any other value in $\mathbb{R}$ \citep{bianchi2010all}.} 

This presentation provides a sense in which the central equation in general relativity is itself defined formally as a matter of convention, in which case the class of vacua in general relativity is itself formally conventional: those spacetimes $(M,g)$ whose geometry is source-free --- that is, associated with vanishing (dynamically relevant) stress-energy --- just according to a conventionally distinguished member of the family. But the fact that, from a mathematical perspective, vacua are defined geometrically as a matter of convention does not preclude that the choice of convention itself be established on further, physical grounds. (The central equation of general relativity is, after all, intended as a geometro\emph{dynamics of gravity}.) As already flagged, the bare cosmological constant is often treated not as conventional, but as a constant of nature with units (length)$^{-2}$. From this perspective, the mathematical convention is evidently to be \emph{decided} with such underlying physics in mind. A broad pluralism about what kinds of physical argument may bear on this decision thus distinguishes between various traditions in the wake of general relativity (with implications for appropriate representations of empty space by spacetimes, or models of the gravitational theory).

For instance, in the tradition I have in mind within this section, the bare term's value is made to vanish, \emph{so as to do away with} the apparent need for any new constant of nature. Einstein's equation is thereby physically necessitated as the distinguished member of the family.\footnote{\label{fnsuspician}The sense in which this traditional commitment corresponds to attitudes of various central figures throughout history is surprisingly tenuous --- again, see \citep{earman2001lambda}. Still, I claim that it is standard enough through that time, fitting naturally with a perpetual suspicion of the geometrization of gravity according to the theory, contrary to the fates of the other known fundamental forces \citep{blum2015reinvention}. In many contexts, general relativity has been regarded --- as far as is possible --- as a classical relativistic field theory just like more familiar others, even in certain restricted applications to be substituted by a classical `spin-2' particle theory of gravity, which is entirely on par with pre-quantum relativistic field theories in particle physics \citep{salimkhani2020dynamical}. Quantizing the spin-2 theory, which takes for granted a vacuum Minkowski spacetime in the background (like do those other field theories in particle physics), leads to an effective field theory of gravitons. This effective field theory, embraced widely in practice, falls under an umbrella that \citet{wallace2022quantum} has termed `low-energy quantum gravity': a theory defined more broadly, by means of a path-integral formulation, so as to allow local graviton scattering calculations within a broader class of spacetime backgrounds and choices of vacua. This broader scope is consonant with a larger portion of fundamental physics applications and practice. But which among the broader class of backgrounds and possible vacua is suitable for any given application of low-energy quantum gravity must then be regarded as an open topic for high-energy quantum gravity research, as in the case of the swampland program in string theory \citep{palti2019swampland,silk2022swampland}.} It follows from the form of Einstein's equation that, in this tradition, vacua are associated with vanishing Ricci curvature: whenever a spacetime metric $g$ (with its associated derivative operator) on a manifold $M$ yields everywhere-vanishing Ricci curvature, the corresponding spacetime $(M,g)$ is --- traditionally --- a vacuum solution to the geometrodynamics of general relativity (and vice versa). Call this decision to make the bare cosmological constant vanish as a matter of physical necessity the `Ricci-flat' tradition for the relativistic geometrodynamics of gravity according to general relativity. 

Whence the normative necessity on display in the Ricci-flat tradition? Methodological conservatism counsels that we may look to the genealogical history of our current understanding of gravity, in order to epistemically secure that vacua be Ricci-flat within the latest theory. As noted in passing by \citet[p. 198]{malament1986newtonian}, the Ricci-flat convention is the unique choice for which Poisson's equation appears in its usual form $R_{ab}=4\pi\rho\xi_a\xi_b$ in the recovery of the non-relativistic theory of geometrized Newtonian gravitation as a limiting case of general relativity, considered the relativistic successor. More generally, for Einstein's equation with cosmological constant $\lambda$, Poisson's equation in the relevant limit is $R_{ab}=(4\pi\rho-\lambda)\xi_a\xi_b$.

Implicit in this technical argument is an assumption that the encompassing geometrized theory in the non-relativistic limit has a Euclidean spatial metric. One might have thought that the influence of the bare cosmological constant in the `non-relativistic' spacetime limit of general relativity --- as understood here --- is really to determine the shape of space: spherical (or maybe elliptic), Euclidean, or hyperbolic. On such a view, it could be natural to read the technical argument as, instead, merely demonstrating the well-known fact that one can always mimic effects of a bare cosmological constant in general relativity by the addition of careful contributions to the dynamically relevant stress-energy, given Einstein's equation --- it simply turns out that the added contributions `survive' in the non-relativistic limit as modifying the total gravitating energy density (a result that is, in fact, entirely expected, just by inspection of the components of the stress-energy tensor in the relativistic theory). But as indicated by \citet[p. 192]{malament1986newtonian}, there is independent reason to demand that the shape of space be Euclidean in the theory made to stand in the non-relativistic (geometrized) limit. Namely, geometrized Newtonian gravitation is itself intended as a geometrization of the more familiar, spatially flat gauge theory of the Newtonian gravitational potential (see also \citep[\S 4.2]{malament2012topics}), given that its successor is a geometrized theory instead of an ordinary field theory. Still, the objection is significant for present purposes. It captures a sense in which any non-vanishing bare cosmological constant term can be considered novel in the general relativistic theory of gravity, when understood as succeeding a non-geometrized Newtonian theory. `Tradition' is well named, in this case: the Ricci-flat tradition settles (by fiat) an ambiguity arising from the geometrization of gravity within general relativity, \emph{precisely by denying that we might help ourselves to something physically new}. 

The Ricci-flat tradition has clear implications for representing empty space, in the wake of general relativity. Vacua, in virtue of being devoid of dynamically relevant stress-energy within the theory, are what can provide suitable spacetime backgrounds in the low-energy study of matter, absent the influence of gravity. So in the wake of general relativity, appropriate spacetime representations of empty space ought to belong to the class of vacua. In the Ricci-flat tradition then, we may conclude that Ricci curvature vanishes in appropriate representations of empty space. 

In fact, \emph{only in} the Ricci-flat tradition does Ricci curvature vanish in appropriate representations of empty space. This hooks up to another point of tradition: only in the Ricci-flat tradition can Minkowski spacetime, the flat spacetime diffeomorphic to $\mathbb{R}^4$ familiar as the kinematic setting of special relativity, appropriately represent empty space as a spacetime in the wake of general relativity. The Ricci-flat tradition thereby preserves something else of an older generation of physics: it allows for special relativity to (continue to) describe relativistic kinematics absent gravity.

Or, rather, it does not preclude that special relativity do so. Nor does it preclude many other relativistic spacetime settings that post-date the advent of general relativity, e.g. gravitational wave spacetimes. And yet, it is common that when ignoring gravity, Minkowski spacetime is singled out as the exclusive target spacetime on which it is appropriate to define local relativistic field theories (including perturbative gravity!).\footnote{As one reviewer notes, Minkowski spacetime is easy to work with, given the affine structure of the tangent space at a point (which is itself familiar as one geometric interpretation of special relativity). Since all metrics approximate the Minkowski metric (indeed, approximate all other metrics) at a point, this suggests a \emph{convenience thesis} to account for the physical tradition discussed presently: just take the mathematically easy spacetime, which locally always approximates spacetime geometry anyway, as the physical model of empty space, suitable as the target spacetime on which to define field theories (and then treat any remaining differences perturbatively). Then, find the geometrodynamics that render vacuum the choice made of convenience, to not cause trouble in interpretation down the road. Convenience surely matters in accounting for aspects of tradition, but this convenience thesis misses the point: we simply do not think of the \emph{gravitational physics} of empty space as a matter of convention, choice to be supplied merely by convenience. Rather, we develop our physical thinking looking out for opportunities to take advantage of any mathematical conveniences, wherever such shortcuts appear physically viable.} This raises a question: is there some strengthening of conditions in the Ricci-flat tradition, which accounts for the elevated status of the special relativistic Minkowski spacetime among the vacua? 

To put the question in another way, is there some spacetime property that physically distinguishes Minkoswki spacetime as the unique `special relativistic' vacuum geometry in general relativity, within the Ricci-flat tradition? If so, then it is in virtue of that spacetime property that Minkowski spacetime is singled out, in the Ricci-flat tradition, as the appropriate representation of empty space in the wake of general relativity.

It turns out that there is a compelling answer to this question provided by the algebraic classifications of pseudo-Riemannian geometries developed by \citet{wolf2011spaces}. Minkowski spacetime is the unique Ricci-flat spacetime that is maximally symmetric, or (equivalently) globally isotropic: families of inertial frames in the tangent spaces of points extend as families of inertial frames in surrounding neighborhoods, including the spacetime as a whole \citep{Belotgoodtimesroll}.\footnote{\label{fnisotropy}More carefully, adopting the definition used by \citet{wolf2011spaces} to secure the relevant results: a spacetime $(M,g_{ab})$ is isotropic (resp. locally isotropic) if, for any Real-valued $r$, the collection of isometries (resp. local isometries) that leave $p\in M$ fixed consistitutes a transitive group on the set of (non-zero) tangent vectors $\xi^a$ at $p$, for which ${g_{ab}}_{|_p}\xi^a\xi^b=r$. So for every spacetime that is (locally) isotropic, its (local) group of isometries has the Lorentz group as a subgroup --- reflecting, at each point, the structure of the identity component of the isometries on the tangent space there. Note that isotropy is stronger than the condition (spatial) isotropy familiar in cosmology textbooks, which only mandates that there be a twist-free foliation of spacetime by Riemannian 3-manifolds that are each isotropic (in the stronger sense). To secure the result summarized in the main text, see \citet[Thm. 11.6.8]{wolf2011spaces}, noting first that, restricting attention to spacetimes, local isotropy already implies constant curvature \citep[Thm. 12.3.1.ii]{wolf2011spaces}, and the only Ricci-flat metrics of constant curvature are the flat metrics. (And for the cases of positive and negative Ricci curvature isotropic vacua, which will be relevent shortly, see Thm. 11.6.7.).} \citet{aldrovandi2007sitter} then provide crucial insight on the physical interpretation of the latter isotropy condition, construed as a global algebraic feature unique to a subset of constant curvature spacetimes. Isotropy ensures the ``quotient character of spacetime'' that is familiar, in special relativity, in terms of rotations, boosts, and reflections in space- or time-orientations about every event. This is the `lightcone structure' of relativity theory, which mandates that the operationally defined `speed of light' through an event be constant, independent of place, time, and (oriented) inertial reference frame. What is forgotten of the full kinematic structure of special relativity in moving to emphasize isotropy is only the set of translation symmetries between distinct events. 

\section{Breaking from tradition}\label{secbreaking}

Such is the Ricci-flat tradition: in the wake of general relativity, empty space in relativistic field theories is uniquely appropriately represented by the special relativistic Minkowski spacetime, distinguished as the only isotropic spacetime amongst the class of vacua picked out by Einstein's equation. Yet, plausibly, measurements of the gravitational dynamics of known sources in a general relativistic regime would support an inference to a non-trivial bare cosmological constant in general relativity.\footnote{Though, for a different perspective on the same upshot, see, e.g., \citep{aldrovandi2007sitter} and \citep{wise2010macdowell}, who essentially favor the embrace of what is, locally, a teleparallel alternative to the general relativistic understanding of non-perturbative classical gravity.} In the context of standard model cosmology, this is one standard way of deflating an inference from observations of accelerated cosmic expansion to the presence of dark energy, in the sense of a new matter field: using known estimates for baryonic and cold dark matter, as well as radiation, one can draw on the simplified version of general relativity provided by the cosmological sector of the theory to empirically constrain the effective cosmological constant parameter $\Lambda$ that appears in the Friedmann equations, which one can then regard as ultimately picking out the Einstein equation with cosmological constant $\Lambda$ as the central equation in the overall theory of general relativity. In other words, on one natural reading of standard model cosmology, the parameter value for the (there aptly named) cosmological constant that I have just written as $\Lambda$, which is relevant in capturing the general relativistic dynamics of large-scale cosmic structure, provides direct empirical access to a fundamental constant of nature --- the bare cosmological constant, or what I previously wrote as $\lambda_0$. 

Adopting this perspective amounts to a break from the Ricci-flat tradition. But this break is nothing mysterious --- it amounts to a change in what exactly we take to be the empirical content of the bare cosmological constant, with a concomitant shift in tradition regarding what kinds of physical argument bear on the choice of geometrodynamics in the gravitational theory general relativity. As a consequence of the break, instead of vacua being those spacetimes associated, by physical necessity, with vanishing Ricci curvature (as insists the Ricci-flat tradition), vacua are now those spacetimes associated with Ricci curvature proportional to the signed value of the measured cosmological constant, e.g. via standard model cosmology. 

What happens, in the new tradition, to our pronouncement on the appropriate representation of empty space by spacetimes, in the wake of general relativity? If the conditions for the latter are that spacetime be isotropic and vacuum, then Minkowski spacetime is appropriate if and only if the measured value of the cosmological constant winds up exactly vanishing.\footnote{\label{fnWallace}One may ask: in the new tradition, is it ever reasonable to expect Minkowski spacetime to persist as an appropriate representation of empty space, in the wake of general relativity? This turns out to be a subtle question, albeit a tangent to the main argument in the article. Earlier versions of this manuscript claimed `no': from a modeling perspective, it is natural to assign measure $0$ to exact values for a Real-valued empirical parameter, whereas an answer `yes' would require that the weight given to one particular such value be non-zero. But David Wallace pushed back: one can (and does!) often imagine justifications for measures that place non-trivial weights on an exactly vanishing empirical cosmological constant parameter. For instance, on one view of the `cosmological constant problem', it is standard to at least entertain a possibility that future physics will reveal a new symmetry that forces the value to be $0$. Suppose now that, as a condition of adequacy on that future physics, the novel symmetry comes to accrue empirical support between now and then. Then such empirical support would also bear on claims to have performed measurements of the bare cosmological constant by means of standard model cosmological modeling. Future us might even be reluctant, except in the face of overwhelming significance, to accept small, non-zero measured values of fit in the cosmological data as anything other than confirmation that the constant exactly vanishes in accordance with theory (indeed, this hypothetical reluctance seems echoed by a significant chunk of the recent history of fundamental physics and cosmology practice in the vicinity). But does that fact about our future selves \emph{presently} bear on our claim to have already performed such measurements? Whether it does --- hence, whether Minkwoski spacetime might persist, in the new tradition, as one possible appropriate representation of empty space --- falls to an open question about how anticipations about future physics, in the sense I develop in \citet{schneiderbetting}, are allowed to factor into current expectations about an empirical modeling parameter in cosmology.} As stated at the beginning of this article, observations since the 1990s support a positive value, instead. The upshot of the break from the Ricci-flat tradition, given such observations that favor a positive value for the cosmological constant, is thus summarized by the following technical result: every isotropic spacetime that is, by the new reckoning, vacuum is isometric to either dS or EdS.\footnote{\label{fnhomothety}In fact, for each positive value of the cosmological constant, every isotropic vacuum spacetime belongs in one or another homothetic equivalence class of spacetimes: the first, for which any one member we might take to define dS; the second, for which any one member we might take to define EdS. But without loss of generality, here I fix the value of the positive cosmological constant. I may then define dS and EdS, respectively, as the members of these respective classes for which the scale factor relating the unique vacuum members therein to dS or EdS is $1$.} 

The first of these --- dS --- is a spacetime familiar to any relativist: one of the earliest in the theory to have been explicitly identified, and which is infamous for its non-trivial cosmological horizons that are defined relative to every possible maximal observer. The second --- EdS --- is less well-known, though it has nearly the same historical legacy (\citet{Belotgoodtimesroll} gives references to work in the 1920s by the mathematician Patrick du Val). One may construct EdS by performing antipodal identifications in dS, treated either as a Lorentzian-signature pseudo-sphere or as an explicitly defined timelike hyperboloid in a five-dimensional Minkowski embedding space. Alternatively, starting from the conformally completed dS, one may identify EdS as the interior to the conformally completed spacetime that is constructed by gluing together past and future conformal boundaries --- identified with an non- orientation preserving `twist' so that the interior remains a spacetime \citep{parikh2003elliptic}.\footnote{\label{fncovering}In four spacetime dimensions, dS is both the time-orientable double cover and universal cover of EdS. For an overview, see \citep{Belotgoodtimesroll}. For more detail and related constructions, see \citep{schrodinger1957expanding} and \citep{calabi1962relativistic}. Finally, for a sense of how to consider relativistic field theories on EdS, given its relationship to dS, see \citep{friedman1997field}.}

From a cosmological perspective, such a conclusion about the representation of empty space --- in terms of either dS or EdS --- is entirely intuitive. A result of including a positive cosmological constant in standard model cosmology is that our large-scale universe ever approaches a far future that is shared by de Sitter spacetime.\footnote{More precisely: our large-scale universe is `future asymptotically de Sitter'. But this is not to say that such a fate is guaranteed \citep{doboszewski2019interpreting}.} This is because the contribution to cosmic expansion provided by the positive-value cosmological constant comes to dominate over the contributions of matter, ultimately dynamically flattening out any inhomogeneities in large-scale cosmic structure across space, e.g. galaxies and clusters, that are due to the attractive character of ordinary gravitational interactions. So too, then, might we like to regard regions within our large-scale universe as, locally, ever approaching vacuum: now understood as a model of cosmic void. And since the large-scale universe is conformally flat in standard model cosmology, the appropriate vacua to approach, on this picture, are only ever those which are locally maximally symmetric, or equivalently locally isotropic. These are vacua of constant curvature, which, of course (in the positive curvature case) are precisely those that are locally equivalent to dS (and so, as well, to EdS).  

From a more global spacetime perspective though, the result is unintuitive. The isotropy of empty space is insufficient to mandate that it have a unique spacetime representation among the new, positive curvature vacua of general relativity. In other words, whereas we previously had a uniqueness result for the representation of empty space in the study of matter from within the Ricci-flat tradition, we now have a non-uniqueness result for the same. In disanalogy to our understanding of special relativitistic geometry as a restriction of general relativity, in the traditional case, to the spacetime setting that ignores gravity (except perhaps by means of local perturbations in the vicinity of an interaction event), now the global spacetime structure of empty space is essentially indeterminate. Empty space is, from a general relativistic perspective, described by an equivalence class consisting of both dS and EdS.\footnote{Similar results by \citet{wolf2011spaces} for isotropic, locally anti-de Sitter spacetimes lead to analogous non-uniqueness arguments for the case of a negative cosmological constant. Hence, in the new tradition whereby the bare cosmological constant is something amenable to direct empirical measurement, it is inevitable (save for the open debate in footnote \ref{fnWallace}) that empty space should wind up described by an equivalence class consisting of more than one spacetime.}

This conclusion suggests that there is no meaningful difference between the spatiotemporal representations provided by dS and EdS, respectively, in our thinking about the low-energy physics of matter, prior to the introduction of a gravitational interaction. But this is quite unusual. dS and EdS register as physically distinct by many standard means of distinction. Most notably, whereas dS is globally hyperbolic --- placing it near the top of what is ordinarily taken to be a ladder (or `hierarchy') of causality conditions in the literature on global spacetime structure  --- EdS fails even to be time-orientable (usually at the bottom-most rung of the ladder; see e.g \citep[ch. 6]{hawking1973large}). Still, what they have in common is that, given any maximal observer in EdS, their causal patch is isometric to the causal patch of a maximal observer in dS, and vice versa  \citep{schrodinger1957expanding}. In fact, the same is true with respect to the observer's causal past and causal future, respectively.\footnote{Here, I assume that the observer in EdS locally carries a time orientation. To make the isometry explicit, one must first extend that orientation all the way up to a horizon--- cf. \citep{hackl2015horizon}. This region of EdS is topologically $\mathbb{R}^4$ and can be covered by a single coordinate patch familiar in cosmology: that of an open, flat spatial expanse, centered on the observer, which exponentially expands (or else contracts) in that observer's proper time. This familiarity is exploited in \citep{aguirre2003inflation}, in the context of inflation.} So the physical difference between dS and EdS is, in some sense, \emph{merely} one of global geometry. 

Perhaps, then, what is happening is that we are being misled --- at least in this particular context --- by standards of physical distinction that are otherwise sacrosanct in relativistic physics.\footnote{In some conversations within the foundations of contemporary physics, non-isometry of spacetimes is taken to be the ultimate standard of physical distinction in general relativity. But as \citet{fletcher2020representational} points out, non-isometric, homethetic spacetimes are often equally adequate in applications of general relativity --- specifically, when global systems of units need not be fixed. Here, I set aside any further considerations about whence the units used to measure geometric properties of a spacetime. The point, rather, is simply that context of application does seem to matter in contemplating standards of physical distinction in classes of spacetimes, as the latter are deployed in the wake of general relativity for various further ends. So, what it is that is sacrosanct may not be quite as succinctly stated as would otherwise be hoped.} Conservatively, in the terms provided by \citet{belot2011geometric}, we might be inclined to say that geometric possibility in our current thinking about a matter-filled universe is, in the absence of gravity, simply not well captured by the class of Lorentzian manifolds, equipped with its canonical standard of inequivalence. 

But this phrasing is somewhat misleading: evidently, spatiotemporal possibility in our current thinking about a matter-filled universe, in the absence of gravity, is simply not well captured by global \emph{geometric} intuitions at all. Rather, empty space is itself a local geometrodynamical and global \emph{algebraic} structure, which ultimately fails to amount to a (global) geometric structure except in the special case of the Ricci-flat tradition, where the bare cosmological constant is made to vanish. Hence, in applications of general relativity to modeling empty space in field theories, we ought to be deflationary about any outstanding global \emph{geometric} differences between dS and EdS --- including such physically loaded differences as time orientability! Interestingly, in the program in dS asymptotics discussed above, the decision to impose boundary conditions along a null surface within dS immediately outside the causal future of an isolated system would seem to reflect this deflationary view --- see especially the discussion in \citep{ashtekar2019asymptotics}. 

\section{Freedom in theorizing}

I have argued so far, with appeal to methodological conservatism, that within the foundations of our contemporary field theories of matter and perturbative gravity, the appropriate spacetime representation of `empty space' is, in the wake of general relativity (as our fully non-perturbative theory of gravity), fully specified as an isotropic vacuum spacetime. As already discussed, in the case of a positive cosmological constant (breaking with the Ricci-flat tradition), this specification yields an equivalence class of geometrizations of empty space, which consists of dS and EdS. The foundational argument thus implies that dS and EdS are both appropriate spacetime representations of empty space --- this, despite what are ordinarily regarded as serious physical differences between them. It strikes me that the identification of this equivalence class, given a positive cosmological constant, does well to contextualize certain speculative projects in holographic quantum cosmology that are otherwise obscure, and which could otherwise even seem to represent instances of radical exploratory departure from our current understanding in physics. Since I think the latter view, especially, is more or less backwards, it is to contextualizing these speculative projects as continuous with our current physics, in light of the foundational argument about empty space, that I should like to dedicate the remainder of the article. 

Holographic quantum cosmology builds on an AdS/CFT conjecture (short for `anti-de Sitter spacetime/conformal field theory', originally within string theory), which associates a perturbative quantum gravitational theory (or semiclassical gravitational theory) within a bulk anti-de Sitter spacetime with a dual, conformal field theory defined along the spacetime's asymptotic boundary.\footnote{\label{fnCTC}Or rather, to avoid closed timelike curves in the background of the conformal field theory along the boundary, one considers as bulk the universal cover of anti-de Sitter spacetime. For a recent general philosophical discussion of AdS/CFT in quantum gravity research outside of the specific context of quantum cosmology, see \citep{jaksland2020holography}. AdS/CFT is of interest in quantum gravity research outside of quantum cosmology because, as one reviewer points out, asymptotically anti-de Sitter settings provide a convenient infrared regulator in physically relevant effective field theories. So, pending AdS/CFT, we may hope to treat physically relevant, gravitationally coupled effective field theories holographically in quantum gravity research.} But in quantum cosmology in light of a positive cosmological constant, some researchers are more interested in a conjectured `dS'/CFT, with the CFT defined along an observer-relative boundary that is motivated by geometric features specific to a dS bulk \citep[\S8  and references therein]{de2016conceptual}. 

Notably, it is then not uncommon to encounter assertions of a certain `freedom' in research on dS/CFT: to \emph{swap out} dS for EdS, moving forward. A passage from \citet[p. 5]{parikh2003elliptic} succinctly captures the move in practice (emphasis added): 
\begin{quote}
``Hence we are using our freedom of topology to \emph{impose} [an antipodal] identification of de Sitter space.''\end{quote}
That physicists are free in this way --- to choose between dS and EdS as just glossed --- is in many respects encouraging: it may turn out to be more convenient --- or ``natural'' \citep[p. 4]{parikh2003elliptic} --- to dwell on one spacetime setting, rather than the other. For instance, whereas dS is globally hyperbolic, the conformal boundary of EdS has only one (timeless!) component. In the context of holographic research in quantum gravity, the latter is arguably a more appealing starting point. But what is the freedom that is being appealed to, in order to choose to start there? In other words: How should we interpret the freedom being asserted?

At a glance, the perspective on display in the quoted passage is a close companion to the view embraced in AdS/CFT, which typically equivocates between AdS and its non-isometric, universal covering spacetime (see footnote \ref{fnCTC}). Indeed, as noted in footnote \ref{fncovering}, dS is the universal cover of EdS, strengthening such an association. This supports reading the invocation of freedom in light of a pragmatic `anything goes' view of theoretical research methodology, along the lines of \citeauthor{feyerabend2010against}'s \citeyearpar{feyerabend2010against} epistemological anarchism. Namely: the asserted freedom in holographic quantum cosmology is, on this take, a radical one. The practicing theorist might just as well consider the prospects of `EdS'/CFT, where others could consider dS/CFT (and still others could consider other formal structures, besides --- for instance, the original AdS/CFT, where dS as a description of the cosmological bulk is demoted to a quasi-local description of some excitation of the AdS ground state). A particular mathematical technique from the study of manifold topology --- universal covers --- is thereby seen as revealing one exploratory route from one curious option to another (or back). But note on this view that the particulars of the mathematical technique are simply not all that important to the physical argument: various other mathematical techniques would just as well illuminate other exploratory routes relating distinct possible new projects.

In a sufficiently general sense, this radical interpretation seems right: theoretical physicists are always free in their professional practice --- in pursuit of variegated research ends --- to tinker about. And it is, perhaps, not so radical an interpretation as to seem trivial: the interpretation proposed here is not that absolutely anything at all goes in equal measure with everything in physics that has come before. It is, for instance, constrained by relevant mathematical tools as have been adopted for use in contemporary spacetime physics --- the theoretical physicists' toolkit for their tinkering.

Still, there is something unsatisfying about the radical interpretation applied to this particular case --- perhaps just that it is insufficiently fine-grained. Radical freedom does not account for why dS serves as starting point, so that a familiar mathematical technique might get us, specifically, to EdS moving forward. Better is an interpretation that makes sense of the specific mathematical argument on offer, not just the kind of mathematics implicated. On this point: I suggest that the construction of EdS from its universal cover is altogether a red herring. Isotropy, given a standard for deciding vacua, is the real culprit: the restriction of focus to just dS or EdS reflects my argument about appropriate spacetime representations of empty space given general relativity, apropos of measurement of a positive cosmological constant. Meanwhile, parting ways from the deflationary attitude raised at the end of the previous section about the difference between dS and EdS, the move now is to treat the global geometry of empty space as some bit of physics that is presently ambiguous --- something like open texture \citep[ch. 1]{hesse2005forces} in current theory, left as-yet unsettled or indeterminable by any rational argument. 

There is some historical precedence for this latter view. \citet{schrodinger1957expanding} suggested that that the difference specifically between dS and EdS is ultimately one of interpretation about observer-relative horizons in one or the other spacetime setting. Namely: one might just as well regard maximal oriented observers in dS with their relative, global cosmological horizons (which screen off whole regions of space and time from the observer) as, instead, maximal observers situated in EdS with relative `orientation horizons' (surfaces at which physically motivated, locally imposed orientations in time simply fail to be further extendable).

In sum: \emph{in light of the non-uniqueness of the spacetime representation of empty space given a positive cosmological constant}, we may interpret the asserted freedom off in this corner of the quantum cosmology literature as something more fine-grained than radical. Sure enough, choosing a holographic approach to quantum cosmology is an expression of the physicist's radical freedom to tinker about. But the further asserted freedom within holography --- to make use of proprietary global structure of EdS over, specifically, that of dS --- is a freedom that may be claimed as won precisely on the basis of the current state of our theoretical scientific knowledge. \emph{What we may claim to know about the spacetime structure of empty space in contemporary physics engenders choice, within holographic quantum cosmology, of exactly this form: dS or else EdS.} It falls to a question about how we might think about observer-relative horizons in our current physics, where it turns out that there is some open texture when focused on the physics of empty space.

\subsection{Speculating about the future}

I have just argued that the freedom being asserted, in the move to employ EdS over dS, is a freedom about how to go about developing some future theory, in light of current understanding of current theory. The decision to swap out dS for EdS is, perhaps, not above critique. But, it is above reproach: our best physics so far simply leaves open how we are to think about the global geometric content of empty space, given a positive cosmological constant, as a description of the observer-relative holographic cosmological bulk. 

A key distinction needed to make this argument is that between the professional physicist, qua decision maker, and the professional physicist, qua epistemic agent. Whereas the epistemic agents are merely content to assert what it is that they know, on the basis of our best physics so far, the decision-making physicists are free, \emph{by the lights of our best physics so far}, to choose how they are to proceed in pursuit of their further aims. In both cases, the physicist is in possession of our current best physical theories to inform their attitudes and actions. But whereas the former is ambivalent between the spacetime representations of empty space provided by dS and EdS given current physics, and maybe even deflationary about the differences between them, the latter may nonetheless see fit to break the tie between the two, in favor of just one. 

Implicit in this view is the notion that a decision-maker's resources may, in some sense, \emph{outstrip} their evidence base. Just so: they have further commitments than merely introspecting on the self-consistent organization of their own, current beliefs. Though it is not stated as such, I take this to be the key insight developed (in a more general context) by \citet{currie2021science}, under the guise of getting clear about speculations in scientific practice. For \citeauthor{currie2021science}, speculations are `function-first', and hence most noteworthy in virtue of their pragmatic bend: what they \emph{do} to/for ongoing scientific research. That is to say, a hypothesis is speculative just when it aims to be productive in some broad sense, given the particular epistemic situation in which that hypothesis is offered. So, good speculations in science are, ultimately, posits that may be justified pragmatically, with respect to the local details and aims of surrounding research. It follows that our embracing a speculation --- whether as a matter of belief or some other doxastic attitude similar to it, cf. \citep{fleisher2018rational} --- is grounded in our confidence in the immediate fruits of our speculating it, in the course of ongoing theorizing.

Note that it is the last qualification --- `in ongoing theorizing' --- that weds criteria for good speculations to a scientist's current evidence base, informed by current theory.\footnote{That is to say, the project of professional scientific research is not one that ever begins anew. Consider: ``Science works through continuity, not discontinuity. [...] building upon a vast and ever growing accumulation of empirical and theoretical knowledge, that provides the hints we need to move ahead''  \citep[p. 489]{rovelli2018physics}.} But it also weds criteria for good speculations to the theorists' forward-looking aims of their research, which reach out beyond current theory. And so, to stress a virtue of the function-first account: note that the epistemic role for speculation need not itself be grounded in a claim of increasing the evidential power of our current theories, augmented by the given speculation. \citeauthor{currie2021science} also allows that speculations can be productive in virtue of opening new areas of research relevant to aims that are possibly decoupled from whatever were the aims in the past (which previously lead to our current theories). For instance, speculations may be good in virtue of providing a scaffold for some further theorizing about an entirely new theory beyond the scope of our current best. 

Taking this view of speculation on board, we may regard the move documented above to employ EdS over dS as one such speculation, justified (so it is claimed) on the upshots that come from exploiting the proprietary global geometric structure of EdS in our ongoing thinking about the appropriate representation of empty space, given a positive cosmological constant. But specifically, those upshots may be understood with respect to new research aims --- e.g. those relevant in quantum gravity research, specifically in holographic quantum cosmology --- that are distinct from others that were previously relevant to fleshing out an appropriate representation of empty space in the context of relativistic field theories.

A worry immediately arises: in what sense are we speculating about \emph{empty space}, i.e. that which we currently understand by the lights of our current physics? \citeauthor{currie2021science}'s account does not provide any obvious leads as to how, in light of the current state of our theoretical scientific knowledge, we are to understand the physical content of such function-first speculations --- however good they are. And recall that, by the lights of the previous section, no argument within our current physics would isolate the employ of EdS over dS with respect to our current relativistic field theoretic physics, in essence because our current best physics fails to distinguish, for such contexts, members of an equivalence class containing both. The upshot is that, by speculating in favor of just one (e.g. EdS), it would seem by the guiding lights of our current best physics that we are simply no longer talking about empty space, per se.

The resolution to this worry, which is essentially a worry about reference, begins with a recognition that any argument in favor of such a speculation as just discussed --- though articulated today --- must depend on expectations about the fate of future theory to come. That is, it is only in the wake of such a future theory that we might eventually find resources, in the context of that future theory, to presently (i.e. \emph{in retrospect}) discriminate between the two choices of representation of empty space for relativistic field theories. Some clarity on this latest subject is, I think, provided by the framework of anticipations I develop and defend in \citep{schneiderbetting} in a different context within quantum gravity research. 

In that framework, one considers theoretical physicists as occupied with a particular kind of task. In their capacities as professional decision-makers, they must leverage their current commitments about the interpretation of our current best theories, such as are constrained by our empirical record, to fashion bets about the future theory they intend to develop. On this account, the theorist who has committed to such bets --- that is, by way of decisions made, in the course of practice --- may inherit demands on their further theorizing, toward the development of that very same future theory. These demands, which are incurred real-time on the basis of decisions made, stipulate some of how the physicists ought to develop the future theory that they are pursuing. Namely, they must proceed \emph{so that} the contents of their bets are borne out, without any subsequent problems, by their eventual embrace of that future theory. These demands take on the status of \emph{anticipations} about the future theory. That is, the physicist anticipates that, by the time an adequate future theory is written down, those particular demands will be satisfied, in effect as a matter of historical necessity (at least, in light of their having taken the bet). 

In fact, I think the relevant bet and anticipation is, in the present case, delightfully easy to spell out. Namely: one bets that the features of EdS are \emph{fundamental} in the study of matter, at low energies and absent gravity, by means of relativistic field theories applied to physics in the observer-relative holographic bulk. Consequently, there is something fundamentally inaccurate about the same such study, which makes use of dS (and likewise, in the case which makes use of both settings, as a matter of indifference). If this is a correct statement of the bet, then the concomitant anticipation is similarly easy to state. In the course of that recovery of current field theoretic characterizations of matter at suitably low energies, details of that recovery will provide resources that single out EdS, rather than dS, as the appropriate representation of empty space. 

One might here ask: what can it mean that, in the course of a theory's recovery, \emph{details of that recovery} will uniquely pick out EdS over dS? After all, if the current theory fails to distinguish between the two, how can we look forward to anything more, specifically from the recovery of that very theory? In fact, a helpful analogy is provided by \citet{weatherall2011some}, concerning the equivalence of gravitational and inertial mass in Newtonian gravitation. As \citeauthor{weatherall2011some} argues, the explanation for the equivalence of gravitational and inertial mass in Newtonian gravitation is \emph{only available} in the wake of general relativity: in context of the Newtonian theory alone, that the two are equivalent is merely an observed regularity, i.e. an empirical fact. But, just as importantly, the explanation is not articulated in terms of general relativity (since, after all, the terms `gravitational mass' and `inertial mass' have no native use in that theory). Ultimately, the explanation of the equivalence crystallizes out of the process of recovering from general relativity Newtonian gravitation, as the predecessor theory, in a suitable limit. The crucial point is that all relevant meanings of terms in the latter theory that are needed to proffer the explanation are implicitly fixed by how one takes that limit.

It is this sense of \emph{intertheoretic, retrospective} explanation, which provides further resources to say something, in the context of Newtonian gravitation as predecessor theory, about the equivalence between the two notions of mass --- even as the successor theory, general relativity, has no obvious use for either such notion. So too, here the thought is that the explanation for why the ambiguity is broken in favor of EdS will come from novel features of the future theory, as those features enter into the recovery of our current field theoretic understanding of matter. Or, at least, that is what is presently being anticipated, when theoretical physicists assert that they are, in their freedom as theorists, betting in favor of EdS in this regime, for use in their ongoing theorizing in holographic quantum cosmology.

There are some disanalogies, of course. Regardless of its ultimate explanation, the equivalence of gravitational and inertial mass in Newtonian gravity remains an empirical fact, in a straightforward sense; the same is obviously not true for the present conversation, concerning the global geometric structure of empty space. So we should not understand the present anticipation as an anticipation that the future theory will be more empirically adequate, on this front (i.e. in the sense of explaining a greater number of empirical facts than its predecessor). This is noteworthy, as discussions about open texture are often conducted in order to better understand the theoretical substrate for progress across theory change, including refining an informal theory at hand \citep{zayton2022open}, such as the ``theory of empty space''. In the case of open texture in physical theory, if progress is tied to an increase in empirical adequacy, the fact that the anticipation on hand is not about increased empirical adequacy suggests that this anticipation is also not about what would make the embrace of the future theory progressive. Learning that empty space, in retrospect, has the global geometric features of EdS would not, by this measure, itself amount to progress.

In a similar spirit, the equivalence of gravitational and inertial mass is an (in retrospect) equivalence between two distinct theoretical constructs relevant in Newtonian gravitation. The present anticipation is that, in retrospect, an equivalence class of theoretical constructs in the current theory may be split apart, in favor of just one of two. So whereas the case of the equivalence of gravitational and inertial mass presents a kind of unification in the move from predecessor theory to successor theory, the present anticipated case does not. Again, on some views of progress in science --- now those favoring unification or local integration --- this could suggest that the present anticipation is not what would make the embrace of the future theory progressive.

Whether accounts of progress in science of either of these kinds ought to be preferred, I do not know. But nor do I think the question is all that important for the present discussion, as good speculations --- as we learn from \citeauthor{currie2021science}'s account --- need not themselves be good in virtue of their direct, ultimate contributions to scientific progress. Theory development is an inevitably complex and human decisions-laden process. In this light, present attitudes about future physics may contribute to the future theory's development in defensible ways. Yet it may still be the case that the ways they contribute wind up entirely orthogonal to assessments about the sense in which what we have learned on the basis of that theoretical research is ultimately progressive.  

\section{Concluding remarks}

There were two parts to this paper: an argument in the foundations of contemporary physics and an argument in philosophy of theoretical physics practice. Regarding the first, I have argued that in the wake of empirical developments in standard model cosmology, the embrace of a positive cosmological constant in general relativity plausibly motivates a view of empty space in our contemporary low-energy study of matter, which is ambivalent between spacetime representations by dS and EdS. This is in contrast with a tradition in which --- paired with an exactly vanishing cosmological constant --- Minkowski spacetime is uniquely favored for the same. The argument provides a basis for dS asymptotics, where the perturbation theory relevant in the study of isolated systems takes for granted asymptotic features associated with dS, rather than those associated with Minkowski spacetime. 

Then I switched topics. I claimed that in light of the foundational argument about empty space, we might turn to consider obscure moves made in practice within one corner of quantum gravity research, concerning speculations about holographic quantum cosmology in the case of a positive cosmological constant. These assertions concerned a freedom to move to exploit, in holographic quantum cosmology, global spacetime structure unique to EdS, specifically as opposed to global spacetime structure unique to dS. I argued that the freedom being asserted in this move to focus on EdS is, in the first place, an expression of the ambivalence between the two global spacetime settings for field theories recovered as low-energy descriptions of the observer-relative cosmological bulk --- i.e. the foundational argument I developed in the first part, applied to holographic quantum cosmology. But then, the argument in favor of the move to EdS is an argument that outstrips current evidence: it is an argument in support of one speculation over others, in the course of ongoing theorizing. I have suggested that such an argument in general can be understood in terms of a function-first account of speculation stated in terms bets, such as result in present anticipations about future theory. Left remaining for study is a question of immediate relevance to ongoing theoretical research, which requires a return to the arena of foundations. Namely: If all I have said thus far is so, is there further any good reason to bet specifically in favor of EdS?

\bibliographystyle{chicago}
\bibliography{dSandEdSFINAL}

\end{document}